\documentclass[pra,showpacs,amsmath,amssymb,superscriptaddress,twocolumn,nofootinbib]{revtex4-1}
\usepackage{graphicx}
\usepackage{dcolumn}
\usepackage{color}
\usepackage{hyperref}

\newcommand*{\figurewidth}{\columnwidth}
\newcommand*{\etal}{\textit{et~al.}}

\def\rma{{\rm a}}
\def\rmc{{\rm c}}
\def\rmr{{\rm r}}
\def\rmH{{\rm H}}

\begin{document}

 \title{Recombination of W$^\mathbf{18+}$ ions with electrons: Absolute rate coefficients from a storage-ring experiment
and from theoretical calculations}

 \author{K. Spruck}
 \affiliation{Institut f\"{u}r Atom- und Molek\"{u}lphysik, Justus-Liebig-Universit\"{a}t Giessen,
Leihgesterner Weg 217, 35392 Giessen, Germany}

 \author{N.~R.~Badnell}
 \affiliation{Department of Physics, University of Strathclyde, 107 Rottenrow, Glasgow G4 0NG, United Kingdom}

 \author{C.~Krantz}
 \affiliation{Max-Planck-Institut f\"{u}r Kernphysik, Saupfercheckweg 1, 69117 Heidelberg, Germany}

 \author{O.~Novotn\'y}
 \affiliation{Columbia Astrophysics Laboratory, Columbia University, 550 West 120th Street, New York, NY 10027, USA}

 \author{A.~Becker}
 \affiliation{Max-Planck-Institut f\"{u}r Kernphysik, Saupfercheckweg 1, 69117 Heidelberg, Germany}

 \author{D.~Bernhardt}
 \affiliation{Institut f\"{u}r Atom- und Molek\"{u}lphysik, Justus-Liebig-Universit\"{a}t Giessen,
Leihgesterner Weg 217, 35392 Giessen, Germany}

 \author{M.~Grieser}
 \affiliation{Max-Planck-Institut f\"{u}r Kernphysik, Saupfercheckweg 1, 69117 Heidelberg, Germany}

 \author{M.~Hahn}
 \affiliation{Columbia Astrophysics Laboratory, Columbia University, 550 West 120th Street, New York, NY 10027, USA}

 \author{R.~Repnow}
 \affiliation{Max-Planck-Institut f\"{u}r Kernphysik, Saupfercheckweg 1, 69117 Heidelberg, Germany}

 \author{D.~W.~Savin}
 \affiliation{Columbia Astrophysics Laboratory, Columbia University, 550 West 120th Street, New York, NY 10027, USA}

 \author{A.~Wolf}
 \affiliation{Max-Planck-Institut f\"{u}r Kernphysik, Saupfercheckweg 1, 69117 Heidelberg, Germany}

 \author{A.~M\"{u}ller}
 \affiliation{Institut f\"{u}r Atom- und Molek\"{u}lphysik, Justus-Liebig-Universit\"{a}t Giessen,
Leihgesterner Weg 217, 35392 Giessen, Germany}

 \author{S.~Schippers}
 \email[Corresponding author, ]{stefan.schippers@physik.uni-giessen.de}
 \affiliation{Institut f\"{u}r Atom- und Molek\"{u}lphysik, Justus-Liebig-Universit\"{a}t Giessen,
Leihgesterner Weg 217, 35392 Giessen, Germany}

\date{\today}

\begin{abstract}
We present new experimentally measured and theoretically calculated rate coefficients for the electron-ion recombination of  W$^{18+}$([Kr]\,$4d^{10}\,4f^{10}$) forming W$^{17+}$. At low electron-ion collision energies, the merged-beam rate coefficient is dominated by strong, mutually overlapping, recombination resonances. In the temperature range where the fractional abundance of W$^{18+}$ is expected to peak in a fusion plasma, the experimentally derived Maxwellian recombination rate coefficient is 5 to 10 times larger than that which is currently recommended for plasma modeling. The complexity of the atomic structure of the open-$4f$-system under study makes the theoretical calculations extremely demanding. Nevertheless, the results of new Breit-Wigner partitioned dielectronic recombination calculations agree reasonably well with the experimental findings. This also gives confidence in the ability of the theory to generate sufficiently accurate atomic data for the plasma modeling of other complex ions.
\end{abstract}

\pacs{34.80.Lx, 34.10.+x, 52.20.Fs}

\maketitle

\section{Introduction}

Tungsten is foreseen as a coating material for plasma facing components in future fusion tokamaks because of its
favorable thermo-mechanical properties. It is the material of choice for the divertor \cite{Pitts2013} of the international ITER tokamak, currently under construction at the Cadarache Research Center in France. Tungsten has already been used successfully in ASDEX Upgrade \cite{Neu2009} and in on-going studies of the ITER-like wall configuration at JET \cite{Matthews2013}. In all these devices it is inevitable that tungsten is sputtered off the inner walls of the vacuum vessel and so contaminates the fusion plasma. Initially neutral tungsten atoms are rapidly ionized via collision processes as they diffuse towards the plasma core. Electron-impact excitation and electron-ion recombination of highly charged tungsten ions lead to subsequent emission of energetic photons which leave the plasma. Above a certain level of tungsten concentration in the core plasma, these radiation losses limit the plasma operation and performance. Plasma model calculations suggest that the fraction of tungsten ions in the core plasma must not exceed a few $10^{-5}$, otherwise plasma burning cannot be sustained \cite{Puetterich2010}. In order to understand the composition of impurities in the plasma, detailed knowledge of the atomic structure of tungsten ions and of the atomic collision processes of tungsten ions in the plasma is required. Thus, excitation, ionization and recombination processes involving tungsten ions are of major interest for the fusion community. Current plasma models for tungsten \cite{Puetterich2008, Puetterich2010} use theoretical recombination rate coefficients from the ADAS database \cite{*[{Atomic~Data~and~Analysis~Structure (ADAS), }] [{}] ADAS} which are based on the semi-empirical Burgess General Formula \cite{Burgess1965}, as discussed in Ref.~\cite{Badnell2003a}.

While investigating tungsten line emission at ASDEX Upgrade, P\"{u}tterich \textit{et al.} \cite{Puetterich2008} had to introduce scaling factors for the ADAS recombination rate coefficients in order to match models of population densities to the observed line intensities. However, good agreement could only be achieved for charge states from W$^{26+}$ and higher. For lower charge states the modeling became increasingly difficult due to the associated large number of spectral lines. The resulting quasi-continuum in the spectrum prevented identification of individual charge states. In order to reproduce the observed line intensities by models, accurate rate coefficients for the dominant excitation, ionization, and recombination processes are needed. Theoretical predictions are challenging because of the complex electronic structure involved. In this situation, experimental recombination rate coefficients are needed to benchmark theory.

To date, only a single direct measurement of a recombination rate coefficient of highly charged tungsten ions has been published, namely for W$^{20+}$([Kr]\,$4d^{10}\,4f^{8}$) forming W$^{19+}$ \cite{Schippers2011}. For this open-$4f$-shell tungsten ion it was found that the recombination rate coefficient is dominated by resonant processes such as dielectronic recombination\footnote{We use the term \lq\lq dielectronic\rq\rq\ recombination to cover all resonant recombination processes since higher-order processes such as \lq\lq trielectronic\rq\rq recombination arise naturally, and are inseparable from the former, in configuration-mixed \lq\lq dielectronic\rq\rq\ recombination calculations.} (DR), in particular at energies below 50~eV, while contributions from radiative recombination (RR) are negligible. The strong, mutually overlapping, low-energy recombination resonances have a significant impact on the total recombination rate coefficient even at the rather high plasma temperatures of interest for fusion devices. A discrepancy of a factor of four was found between the experimental results and the ADAS recombination rate coefficient.

Subsequent to the measurement for W$^{20+}$, new theoretical calculations of recombination rate coefficients of Xe-like tungsten have been carried out. The theoretical calculations have been challenged by the extraordinary complexity of the open-$4f$-shell atomic structure of W$^{20+}$. For such complex systems, the common approach of including correlations via large configuration interaction expansions cannot be applied to the extent that would be necessary to obtain results with sufficient  accuracy. Consequently, intermediate coupling (IC) calculations \cite{Badnell2012} result in smaller resonance strengths than the measured ones at low collision energies.

While the cause of this discrepancy is well understood now, it is technically hard to overcome. In this situation, statistical theory \cite{Flambaum1994,Flambaum2002} provides a useful framework for estimating the \lq\lq missing\rq\rq\ recombination resonance strength. The application of statistical theory to describe the highly-mixed dielectronic capture/autoionization processes via a Breit-Wigner redistribution leads to much better agreement with the experimental merged-beam rate coefficient for W$^{20+}$ \cite{Badnell2012,Dzuba2012}, at least at very low energies. At higher energies, autoionization into excited states becomes energetically possible and this greatly suppresses, or damps, the DR rate coefficient, as was evidenced by the IC results in Ref.~\cite{Badnell2012}. However, the simple statistical model used in Refs.~\cite{Badnell2012,Dzuba2012} did not allow for such damping and so at higher energies the statistical model rate coefficients were shown \cite{Badnell2012} to be much larger than both the (damped) IC results and the experimental results. The recent work of Dzuba \etal~\cite{Dzuba2013} included damping in their statistical approach and they obtained a better, consistent description of the fall-off of the  measured W$^{20+}$ recombination rate coefficient towards higher energies. In the present work, we allow for damping in both our IC calculations (as usual) and in our Breit-Wigner partitioned DR calculations.

In this paper, we present absolute experimental and theoretical rate coefficients for electron-ion recombination of W$^{18+}$([Kr]\,$4d^{10}\,4f^{10}$) forming W$^{17+}$. Experimental rate coefficients were obtained by storage-ring measurements employing the merged-beam technique \cite{Phaneuf1999} at a  heavy-ion storage ring. Experimental details can be found in
Sec.~\ref{sec:exp}. A description of the theoretical calculations is given in Sec.~\ref{sec:theo}. Results are described and discussed in Sec.~\ref{sec:res}. A summary and conclusions are given in Sec.~\ref{sec:con}.

\section{Experiment and excited state population} \label{sec:exp}

The present measurements were performed at the TSR heavy-ion storage ring \cite{Grieser2012} of the
Max Planck Institute for Nuclear Physics in Heidelberg, Germany. The experimental procedures and data analysis  are very similar to the ones used in our previous study on W$^{20+}$ ions~\cite{Schippers2011}. W$^{18+}$ ions were produced by stripping of a parent beam of negatively charged tungsten carbide that was created in an ion sputter source delivering currents
of about 12~$\mu$A. The WC$^-$ ions were injected into a tandem accelerator where carbon atoms and electrons were stripped off by passing the beam through thin carbon foils. Behind the acceleration section isotopically pure $^{182}$W$^{18+}$ ions were selected using a dipole magnet and subsequently injected into the storage ring. The time-averaged electrical current behind the analyzing magnet was 250~pA. The kinetic energy of the stored ions was 169~MeV, corresponding to a velocity of 4.5\% of the speed of light.

The TSR electron cooler was used for electron cooling of the stored W$^{18+}$ ion beam and as an electron target for the present recombination measurements. The recombined W$^{17+}$ ions were separated from the stored W$^{18+}$ beam in the TSR bending magnet following the cooler. The recombination products were detected by a channeltron-based single particle detector \cite{Rinn1982} with practically 100\% detection efficiency. Count rates of up to several tens of kHz were recorded. At these count rates dead time effects were negligible since the detection system can process count rates of up to several hundreds of kHz.

At the beginning of each measurement cycle W$^{18+}$ ions were injected into the storage ring and first cooled for 1.5~s with the cooler cathode voltage adjusted for matching electron and ion velocities. The 1.5~s cooling time also allowed for the de-excitation of metastable W$^{18+}$ ions that are produced in the foil-stripping process. For an estimation of the remaining metastable fraction in the cooled-ion beam, lifetimes of metastable levels of the W$^{18+}$ ground configuration [Kr]$\,4d^{10}\,4f^{10}$ and of the first excited configurations [Kr]$\,4d^{10}\,4f^{9}\,5s$ and [Kr]$\,4d^{10}\,4f^{9}\,5p$ were calculated employing the \textsc{Autostructure} atomic structure code (see Sec.~\ref{sec:theo}). In this calculation, the ground level is found to be [Kr]$\,4d^{10}\,4f^{10}\;^{5\!}I_8$ as was predicted earlier \cite{Kramida2009}. In addition, there are 1670 excited levels within the chosen set of  electron configurations. Their excitation energies range up to about 114~eV above the ground level. Their lifetimes were determined by calculating E1, M1, and E2 radiative transition rates to all accessible energetically lower states. The results for all levels with lifetimes longer than 10~ms can be  found in Tab.~\ref{tab:meta}.

\begin{figure}[b]
\centerline{\includegraphics[width=\figurewidth]{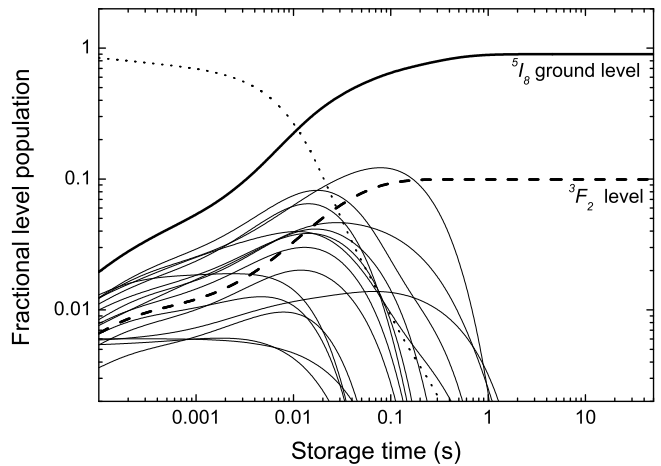}}
\caption{\label{fig:meta}Populations of the 1671 levels of the [Kr]$\,4d^{10}\,4f^{10}$ ground configuration and the [Kr]$\,4d^{10}\,4f^{9}\,5s$ and [Kr]$\,4d^{10}\,4f^{9}\,5p$ first excited configurations of W$^{18+}$ as a function of ion storage time. The thick solid line represents the population of the [Kr]$\,4d^{10}\,4f^{10}\;^5I_8$ ground level, the dashed line denotes the population of the long-lived metastable [Kr]$\,4d^{10}\,4f^{10}\;^{3\!}F_2$ level. The thin solid lines represent the remaining 17 levels from Tab.~\ref{tab:meta}. The dotted line represents the sum of the populations of the 1652 short-lived levels, which are not listed in Tab.~\ref{tab:meta}.}
\end{figure}

Except for the [Kr]$\,4d^{10}\,4f^{10}\;^{3\!}F_2$ level that has a radiative lifetime of about 12 years, all of the calculated lifetimes are below one second. All calculated transition rates were used to simulate the level populations in the stored W$^{18+}$ beam  as a function of storage time. To this end, a set of coupled rate equations \cite{Lestinsky2012} has been solved numerically. As an initial condition, a Maxwell-Boltzmann distribution of the levels  \cite{Lestinsky2012} has been assumed. Figure \ref{fig:meta} shows the resulting populations as a function of storage time. After 1~s about 90\% of the stored ions have decayed to the ground level and most of the remaining 10\% have accumulated in the long-lived metastable [Kr]$\,4d^{10}\,4f^{10}\;^{3\!}F_2$ level. This result is largely independent of the temperature that characterized the Boltzmann distribution of initial level populations. Thus, we conclude that, after the initial cooling of the ion beam, 90\% of the stored W$^{18+}$ ions were in the [Kr]$\,4d^{10}\,4f^{10}\;^{5\!}I_8$ ground level and 10\% remained in the [Kr]$\,4d^{10}\,4f^{10}\;^{3\!}F_2$ level. Because of the very long lifetime of this level, this beam composition did not change during the measurement time interval that followed the 1.5~s cooling period.

\begin{table}
\caption{\label{tab:meta}W$^{18+}$ levels from the [Kr]$4\,d^{10}\,4f^{10}$, [Kr]$\,4d^{10}\,4f^{9}\,5s$, and [Kr]$\,4d^{10}\,4f^{9}\,5p$ configurations with calculated lifetimes longer than 10~ms. $E_\mathrm{ex}$ is the excitation energy from the [Kr]$4\,d^{10}\,4f^{10}\;^{5\!}I_8$ ground level.  Numbers in brackets denote powers of ten.}
\begin{ruledtabular}
\begin{tabular}{@{}dld}
		\multicolumn{1}{c}{$E_\mathrm{ex}$ (eV)} & level & 		\multicolumn{1}{c}{lifetime (s)}\\
    \hline\rule[0mm]{0mm}{12pt}
   0    &    $4f^{10}\;^{5\!}I_8$	   &   	       		\multicolumn{1}{c}{$\infty$}  \\
  2.977 &	 $4f^{10}\;^{5\!}I_6$	   &	      2.13[-2]  \\
  3.543 &    $4f^{10}\;^{5\!}I_5$	   &  	      2.22[-1]  \\
  4.273 &	 $4f^{10}\;^{5\!}F_5$	   &	 	  2.55[-2]  \\
  4.390 &	 $4f^{10}\;^{5\!}I_4$	   &	 	  7.19[-2]  \\
  4.650 &	 $4f^{10}\;^{3\!}F_2$	   &	 	  3.79[+8] \\
  5.271 &	 $4f^{10}\;^{5\!}F_4$	   &	 	  3.80[-2]  \\
  5.982 &	 $4f^{10}\;^{5\!}F_3$	   &	 	  4.28[-2]  \\
  5.862 &	 $4f^{10}\;^{5\!}S_2$	   &	 	  4.13[-2]  \\
  6.331 &	 $4f^{10}\;^{5\!}F_1$	   &	 	  1.72[-2]  \\
  7.090 &	 $4f^{10}\;^{3\!}L_9$	   &	 	  3.41[-2]  \\
  8.032 &	 $4f^{10}\;^{5\!}G_3$	   &	 	  1.29[-2]  \\
  8.049 &	 $4f^{10}\;^{3\!}K_6$	   &	 	  1.16[-2]  \\
  8.498 &    $4f^{10}\;^{3\!}M_{10}$   &    	  0.23[-1]  \\
  9.542 &    $4f^{10}\;^{3\!}P_0$	   &  	      1.97[-2]  \\
 12.050 &    $4f^{10}\;^{5\!}D_2$	   &  	      1.21[-2]  \\
 19.775 &    $4f^{\,9}\,5s\;^{5\!}M_{11}$&        6.29[-1]  \\
 19.988 &    $4f^{\,9}\,5s\;^{5\!}M_{10}$  &      1.74[-2]  \\
 26.484 &    $4f^{\,9}\,5s\;^{3\!}O_{12}$  &      3.71[-2]  \\
\end{tabular}
\end{ruledtabular}
\end{table}

Dielectronic recombination from excited levels is  normally strongly suppressed at all energies compared to that from the ground level. This is due to autoionization into the continuum of levels which lie below the initial metastable one. Consequently, to a good approximation, the experimental cross sections can be multiplied by a correction factor $f_\mathrm{corr}=1.1$ to take account of the 10\% fractional population of the [Kr]$\,4d^{10}\,4f^{10}\;^{3\!}F_2$ metastable level.

For the measurement of the W$^{18+}$ recombination rate coefficient the cathode voltage was ramped through a preselected range of values corresponding to the desired collision energy interval. Each voltage range comprised 2000 discrete collision energy steps. The dwell time was 1~ms at each step, resulting in an overall ramping time of 2~s. Fresh ions were injected into the storage ring and cooled for 1.5~s prior to the next ramping cycle. This scheme was repeated for usually about 1~h,
then the energy range of interest was changed to the next interval. Each scan over a certain energy range had 50\% overlap with the previous measurement. In total, the present measurements comprise collision energies ranging from 0.2~meV to 300~eV.

The experimental energy spread is determined by the velocity distributions of the ions and of the cooler electron beam. It can be characterized by the longitudinal and transverse temperatures $k_BT_\|$ and $k_BT_\perp$ \cite{Kilgus1992}. For a well-cooled ion beam, the velocity distribution of the ions can be neglected and the experimental energy spread is determined by the electron beam temperatures only. In the present experiment the ion beam is only cooled for 1.5~s after injection and there is no beam cooling during the ramping cycles. Therefore, the collision velocity spread, and, hence, the effective temperatures are higher than with the usual experimental scheme (see, e.g., \cite{Schippers2001c}) where beam cooling is applied in between two cooler cathode voltage steps. From the comparison between our theoretical calculations and our experimental measurements (see below) we infer $k_BT_\|\approx0.2$~meV and $k_BT_\perp\approx20$~meV as rough estimates. With these
temperatures the experimental energy spread \cite{Mueller1999c} is 0.05~eV at an energy of 1~eV and 0.80~eV at 290~eV.

For the present measurements, no dedicated effort has been made to calibrate the experimental energy scale beyond the accuracy that is determined by the merged-beam experiment itself. The velocity-matching condition, corresponding to vanishing collision energy of electrons and ions and referred to as the 0~eV case, is found by observing the cusp in the rate at the
recombination detector as a function of the electron acceleration voltage. The acceleration voltage difference to this 0~eV structure defines the experimental electron-ion collision
energy \cite{Kilgus1992}. Its systematic uncertainty lies at sub-meV values near 0 eV and increases with increasing energy. A conservative
estimate \cite{Kilgus1992} yields systematic uncertainties of 0.3 and 1.2~eV at electron-ion collision energies of 10
and 300~eV, respectively.

\subsection{Relative merged-beam recombination rate coefficient} \label{sec:analysis}

From the signal count rate $R$ registered by the recombination detector, the merged-beam recombination rate coefficient as a function of collision energy $E_\mathrm{col}$ is derived as \cite{Bernhardt2011a}
\begin{equation}\label{eq:rel}
	\alpha(E_{\mathrm{col}}) = \frac{R(E_{\mathrm{col}})f_\mathrm{corr}}
       {(1-\beta_i \beta_e)\ \epsilon\ N_i \ n_e(E_{\mathrm{col}})\ L_{\mathrm{eff}}/C}.
\end{equation}
Here, $\beta_i$ and $\beta_e$ are the ion and electron velocities, respectively, in the laboratory frame of reference in units of the speed of light, $\epsilon = 0.97$ is the detection efficiency, $N_i$ is the number of stored ions, $n_e$ is the electron density in the interaction region, and $C = 55.4$~m is the TSR closed orbit circumference.

The effective length $L_\mathrm{eff}$ of the interaction region is different from the length $L = 1.5~$~m of the cooler, because the velocity vectors of electrons and ions point into different directions in the toroidal merging and demerging sections of the cooler. This shortens the length of the merging section, where electrons and ions move with the preset relative velocity; and in the toroidal sections it introduces  higher electron-ion collision energies than the nominal set value. This affects the measured merged-beam rate coefficient, in particular, in energy ranges where it exhibits steep gradients. In principle, this effect can be accounted for by a deconvolution procedure \cite{Lampert1996}. However, this procedure requires knowledge of the electron-ion recombination rate coefficient at higher energies, which is presently not available. Therefore, we have chosen $L_\mathrm{eff}=1.4\pm0.1$~m as the mean value of the geometrically shortest (1.3~m, excluding the toroidal sections) and longest (1.5~m, including toroid sections) overlap lengths, with the uncertainty being half the difference between these two values.

Usually, the number $N_i$ of stored ions is derived from the measured ion current in the storage ring. However, under the present experimental conditions the ion current was too low to be measured using the TSR ion current transformer. Therefore, in a first step, a relative recombination rate coefficient was obtained by normalization of the measured recombination count rate to a proxy of the ion current. In a second step, detailed below, the resulting relative recombination rate coefficient was scaled to the separately measured absolute rate coefficient at zero electron-ion collision energy. The ion current proxy was obtained from the count rate of W$^{19+}$ ions, resulting from ionization in residual gas collisions, on an appropriately situated detector similar to the one used to record the recombination signal. The measurement energy range was well below the ionization threshold of W$^{18+}$ at 462.1~eV \cite{Kramida2006}. Therefore, the ionization signal only depends on the parent ion current and the density of the residual gas, which is assumed to be constant in the relevant part of the TSR for the duration of the data taking.

The relative recombination rate coefficient from Eq.~(\ref{eq:rel}) contains a background that results from electron capture during collisions of the W$^{18+}$ primary ions with residual gas particles. Usually, this background is measured by inserting interleaving reference energy steps into the sequence of measurement energies (see, e.g., \cite{Schippers2011}). However, this procedure significantly reduces the duty cycle of the measurement procedure. In view of the extremely short beam lifetime of only 1.6~s (see below) no interleaving reference steps were used for the present measurements. Instead, we assume that the recombination background from collisions is independent of the electron-ion collision energy and take as a background the lowest measured recombination count-rate level which was measured at an electron-ion collision energy of $\sim\;$260~eV.

After this background subtraction, the relative recombination rate coefficient, Eq.~(\ref{eq:rel}), is put on an absolute scale as described in Sec.~\ref{sec:abs}. With this normalization, the absolute rate coefficient at low energy is found to range up to $>10^{-6}$~cm$^3$~s$^{-1}$ (see Sec.~\ref{sec:resmb}). At  energies above 220~eV its value becomes smaller than $3\times10^{-10}$~cm$^3$~$s^{-1}$ and monotonically further decreases up to $\sim$ 250~eV. Nevertheless, the measured signal at 260~eV can still contain contributions from electron-ion recombination events that have occurred in the cooler. These would be falsely subtracted in the background removal described above. In order to account for at least part of this signal we re-added, after background subtraction and proper absolute normalization (see below), a theoretical rate coefficient for radiative recombination (cf.,~Sec.~\ref{sec:theo}). It should be noted that both the residual variation of $\alpha(E_\mathrm{col})$ above 220~eV and the re-added radiative recombination rate coefficient ($\sim\;2\times 10^{-12}$~cm$^3$~s$^{-1}$ at 260 eV) represent only small corrections to the total rate coefficient.

The major uncertainty associated with the present background correction procedure comes from the neglect of unresolved recombination resonances which also may contribute to the measured recombination signal at 260~eV. If such resonances were present, too much background would have been subtracted, and our experimental rate coefficient would be too small. However, our theoretical calculations do not suggest strong recombination resonances at electron-ion collision energies
around 260~eV (see below).

\subsection{Absolute recombination rate coefficient} \label{sec:abs}

\begin{figure}[b]
\centerline{\includegraphics[width=\figurewidth]{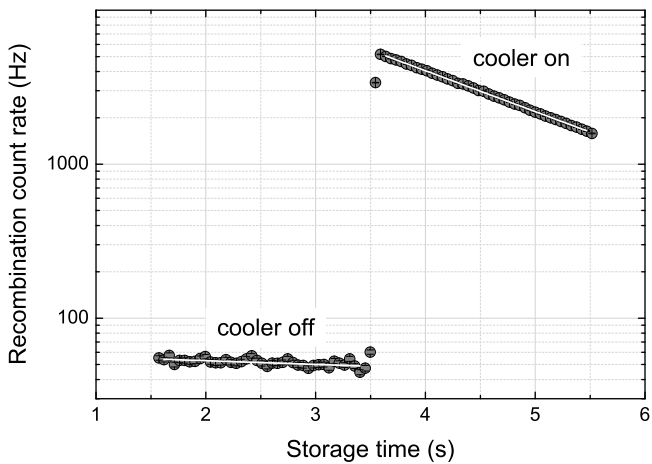}}
\caption{\label{fig:life}Lifetime measurements of the stored ion beam with the cooler electron-beam off and on,
respectively. After 3.5~s the electron-beam was switched on. The symbols represent the measured count rate on the
recombination detector. The white solid lines are exponential decay fits to these data points.}
\end{figure}

As in our previous study with W$^{20+}$ ions \cite{Schippers2011}, the absolute recombination rate coefficient $\alpha_0$ at a collision energy of 0~eV was determined by monitoring the storage lifetime of the W$^{18+}$ ion beam. To this end, the count rate of the recombined W$^{17+}$ has been recorded as a function of beam storage time. The lifetime of the ion beam is limited by collisions with residual gas particles. Due to additional electron-ion recombination, the lifetime is even further reduced when the electron beam of the cooler is switched on. The measured count rates over time, with the cooler switched on and off, were fitted with separate exponential decay functions
(Fig.~\ref{fig:life}). The absolute recombination rate coefficient can be determined from the respective beam lifetimes $\tau_\mathrm{on}$ and $\tau_\mathrm{off}$ obtained from the fits via \cite{Pedersen2005, Novotny2012}
\begin{equation}\label{eq:abs}
	\alpha_0 = \frac{\tau_{\mathrm{on}}^{-1} - \tau_{\mathrm{off}}^{-1}}{n_e L_{\mathrm{eff}}/C}.
\end{equation}
The electron density at zero electron-ion collision energy was $n_e = (10.0\pm0.1)\times 10^{6}$~cm$^3$. The beam lifetimes  $\tau_\mathrm{on}= 1.62\pm 0.02$~s and $\tau_\mathrm{off} = 14\pm 4$~s. These values were obtained by averaging over the fit results from three separate measurements and result in $\alpha_0 = (2.16 \pm 0.09) \times 10^{-6}$~cm$^3$~s$^{-1}$.
The separate fit results from each individual measurement agreed within the uncertainties from the fit. The quoted uncertainties correspond to a 90\% confidence interval. This absolute recombination rate coefficient at 0~eV collision energy was then used to normalize the relative merged-beam recombination rate coefficient which was obtained by scanning the collision energy as described above.

It should be noted that the energy-independent factor $f_\mathrm{corr}$ from Eq.~(\ref{eq:rel}),
that accounts for the metastable ion fraction in the parent ion beam, effectively does not enter the absolute normalization of the cross section via Eq.~(\ref{eq:abs}). In principle, one could expect different beam lifetimes for ground state ions and metastable ions. This would lead to double-exponential decays for each part of Fig.~\ref{fig:life}. However, the observed beam decays in Fig.~\ref{fig:life} are both single-exponential. There are two possible explanations. First, the long-lived [Kr]$4\,d^{10}\,4f^{10}\;^{3\!}F_2$ level is not significantly populated. Second, the relevant collision cross sections are nearly the same for both the [Kr]$4\,d^{10}\,4f^{10}\;^{3\!}F_2$ metastable level and the [Kr]$4\,d^{10}\,4f^{10}\;^{5\!}I_8$ ground level. Consequently, the decay curves do not allow one to discriminate between the two levels and the derived value for $\alpha_0$ is independent of the population of the metastable level. In either case, $\alpha_0$ is the correct value for the recombination rate coefficient of ground-level ions and there is no additional uncertainty of this value related to $f_\mathrm{corr}$.

At a confidence limit of 90\%, the statistical error of the absolute rate coefficient at zero collision energy amounts to 4.2\%. Systematic uncertainties of the absolute rate coefficient arise from several sources. The systematic uncertainty of the effective interaction length amounts to 7\% and that of the electron density to 1\% \cite{Kenntner1995}. The systematic error from background subtraction depends on the collision energy. At 0 eV, where the recombination rate coefficient is independently measured via Eq.\ (2), there is no influence of the background subtraction at all. At high collision energies of 220~eV, where the residual recombination signal after the background subtraction is small, the resulting uncertainty amounts to $\sim\;$80\%. At intermediate energies of 1~eV and 30~eV, the background subtraction procedure results in systematic uncertainties of 2\% and 25\%, respectively. Since all these uncertainties are independent of each other, they need to be summed in quadrature. In addition to the systematic uncertainty there is a counting-statistical error on the relative recombination rate coefficient (as displayed in Fig.~\ref{fig:exptheo}), which varies with energy as well. The total uncertainty of the data at a 90\% confidence limit, i.e., the quadrature sum of systematic and statistical uncertainty, ranges from 8\% at 0 eV across
9\% at 1 eV, 38\% at 30 eV and 120\% at 220 eV, as the rate coefficient approaches zero. Different errors are derived for the plasma rate coefficient as detailed below.

\section{Theory} \label{sec:theo}

Our basic approach to dielectronic recombination is detailed in \cite{Badnell2006b}.
We use the independent processes, isolated resonances plus distorted waves (IPIRDW) approximation.
We energy-average each resonance over a width of energy $\Delta E$ which is chosen to be
large compared to the resonance width and small compared to the characteristic width of
any subsequent convolution. The choice of $\Delta E$ is arbitrary and is usually taken
to be a constant (linear or logarithmic).

Let $\bar{\sigma}^{j}_{f\nu}(E_\rmc)$ denote the partial energy-averaged dielectronic recombination
cross section, centered on $E_\rmc$, from an initial state $\nu$ of an ion $X^{+z}$,
through an autoionizing state $j$, into a resolved final state $f$ of an ion $X^{+z-1}$. Then
\begin{eqnarray}
\bar{\sigma}^{j}_{f\nu}(E_\rmc)&=&{\left(2\pi a_0 I_{\rmH}\right)^{2} \over \Delta E \,\, E_\rmc }
{\omega_{j}
\over 2\omega_{\nu}}
\nonumber \\
 &\times&
{ \tau_0\sum_{l}A^{{\rma}}_{j \rightarrow \nu, E_{\rmc}l} \, A^{{\rmr}}_{j \rightarrow f}
\over \sum_{h} A^{{\rmr}}_{j \rightarrow h} + \sum_{m,l} A^{{\rma}}_{j \rightarrow m, E_{\rmc}l}}\,,
\label{pdri}
\end{eqnarray}
where $\omega_j$ is the statistical weight of the
$(N+1)$-electron doubly-excited resonance state $j$, $\omega_\nu$ is the statistical weight
of the $N$-electron target state (so, $z=Z-N$, where $Z$ is the nuclear charge) and the autoionization
($A^{\rma}$) and radiative
($A^{\rmr}$) rates are in inverse seconds. Here, $E_{\rmc}$ is the energy of the continuum
electron (with orbital angular momentum $l$), which is fixed by the position of the resonance $j$
relative to the continuum $\nu$,
$I_{\rmH}$ is the ionization potential energy of the hydrogen atom (both in the
same units of energy)
and $(2\pi a_0)^{2}\tau_0=2.6741\times 10^{-32}$~cm$^2$s.

We usually sum over all resonances $j$ so as to compare with experiment or for application
to plasma modeling. It is convenient to \lq\lq bin\rq\rq\ the cross section via
\begin{eqnarray}
\label{ebin}
\bar{\sigma}_{\nu}(E_n)=\sum_{j} \bar{\sigma}^{j}_{\nu}(E_\rmc)
                       &=&\sum_{j,f} \bar{\sigma}^{j}_{f\nu}(E_\rmc)\\
&\forall& E_\rmc \in [E_n, E_{n+1})\,,\nonumber
\end{eqnarray}
where $E_{n+1}=E_n+\Delta E$ (for the linear case). The sum over $f$ is over all final states
which lie below the ionization limit of the recombined ion $X^{+z-1}$.
This sum may include cascade through autoionizing levels
in general although we do not need to consider it here.
The sums over $f$ and $j$ are taken to convergence to obtain total rate coefficients
for application to low-density plasmas but the sum over $f$ (and hence $j$)
normally needs to be truncated for application to laboratory measurements.

Our calculational approach closely follows that used for W$^{20+}$ \cite{Badnell2012},
with one extension. We used the program {\sc autostructure} \cite{Badnell2011} to calculate all energy levels,
radiative rates and autoionization rates necessary to describe the full range of two-step DR reactions
which take place via $\Delta n=0$ and $\Delta n=1$ promotions of $4d$ and $4f$ electrons from the
W$^{18+}$ ground-state. We used configuration-average-, $LS$-~, and intermediate-coupling schemes.

The purpose of using multiple coupling schemes is to study the convergence of theory with
experiment at low energies as the amount of mixing of autoionizing states is increased ---
see Fig.~5 of \cite{Badnell2012}. Even the intermediate coupling results fall short of
experiment because we are restricted to mixing autoionizing states which result from
one-electron promotions (plus capture). There are many more autoionizing states present
which result from multiple-electron promotions (plus capture). These are not populated directly by
dielectronic capture from the ground state since this is mediated by a two-body operator.
Nevertheless, such \lq\lq forbidden\rq\rq\ capture states could typically radiatively stabilize
at a rate $A^\rmr$ comparable with that for an \lq\lq allowed\rq\rq\ capture, if they were populated somehow.
Such population occurs through mixing of doubly excited states with-and-between multiply excited states.

A simple model~\cite{Badnell2012}. If the autoionization rates $A^\rma$ corresponding to the allowed dielectronic captures
(i.e., in the numerator of Eq.~(\ref{pdri})) initially satisfy
\begin{eqnarray}
A^\rma\ll A^\rmr
\end{eqnarray}
then (see Eq.~(\ref{ebin}) also)
\begin{eqnarray}
\sum_j\bar{\sigma}^{j}_{\nu}\propto A^\rma
\end{eqnarray}
both with-and-without mixing (provided $A^\rma\ll A^\rmr$ in the denominator as well).
Thus, the $\bar{\sigma}^{j}_{\nu}$
are merely redistributed by the unitary mixing transformation acting on states $j$.

However, if initially
\begin{eqnarray}
A^\rma\gg A^\rmr
\end{eqnarray}
then
\begin{eqnarray}
\sum_j\bar{\sigma}^{j}_{\nu}\propto A^\rmr\,.
\end{eqnarray}
But, following complete redistributive mixing of $A^\rma$, such that $A^\rma\ll A^\rmr$ again,
we have
\begin{eqnarray}
\sum_j\bar{\sigma}^{j}_{\nu}\propto A^\rma\,,
\end{eqnarray}
i.e., enhanced by a factor $A^\rma/A^\rmr$ compared to the unmixed result.

The open $f$-shell is a situation where such redistributive mixing occurs. For example, for W$^{20+}$($4f^8$)
a factor of three enhancement of the low-energy DR cross section was found \cite{Badnell2012, Dzuba2012}
compared to the standard  intermediate coupling results. Indeed, Gribakin and Sahoo
\cite{Gribakin2003a} have demonstrated the chaotic nature of the mixing for the DR of Au$^{25+}$($4f^8$).
However, it should be noted that as the $f$-shell closes-off then the DR measurement \cite{Schippers2011a}
for Au$^{20+}$($4f^{13}$) is well described conventionally~\cite{Ballance2012}.
Statistical theory~\cite{Flambaum1994} as applied to DR~\cite{Flambaum2002} essentially
reduces to the usual sub-configuration average representation for DR but with a Breit-Wigner weighted redistribution
of the dielectronic capture/autoionization --- in particular, compare Eq.~(5) of \cite{Dzuba2013} with Eq.~(5)
of \cite{Pindzola1991a}.
Dzuba et al \cite{Dzuba2012, Dzuba2013} redistribute explicitly over multiply-excited sub-configurations
whilst we partition them uniformly over arbitrary bin widths assuming a quasi-continuum of levels~\cite{Badnell2012}.

We define a new set of autoionizing levels $\bar{j}$ to be used in Eqs~(\ref{pdri}) and (\ref{ebin})
in place of $j$. The autoionization rates as a function of $j$ are redistributed over $\bar{j}$ via
\begin{eqnarray}
A^{{\rma}}_{\bar{j} \rightarrow \nu, E_{\bar{\rmc}}l} &\leftarrow&
A^{{\rma}}_{j \rightarrow \nu, E_{\rmc}l}L_{\bar{j}}(E_{\rmc})
\label{red}
\end{eqnarray}
where the Breit-Wigner weighting $L_{\bar{j}}$ is given by
\begin{eqnarray}
L_{\bar{j}}(E_{\rmc})&=&{\Gamma/(2\pi) \over \left(E_{\bar{j}}+E_\nu -E_{\rmc}\right)^2 + \Gamma^2/4}\,,
\label{lor}
\end{eqnarray}
$E_{\bar{\rmc}}=E_{\bar{j}}+E_\nu$,
and
$\Gamma$ is the spreading width for the redistribution which characterizes the chaotic mixing in
the open $f$-shell.
The results are not sensitive to the precise value of this width
since we are in the complete redistributive regime
and we use the same value as for $W^{20+}$~\cite{Badnell2012} viz.\ 10~eV, as suggested by
large-scale structure calculations~\cite{Flambaum2002}.
The choice of $\bar{j}$ is essentially arbitrary when the fluorescence yield of Eq.~(\ref{pdri})
is taken to be unity. For example, we can define (partition) $\bar{j}$ by our bin energies (\ref{ebin})
viz. $E_{\bar{j}}=E_n-E_\nu$. Note: since each redistributed resonance is partitioned over many bins
only $\int^{n+1}_{n}L_{\bar{j}}(E)dE\approx L_{\bar{j}}(E_{\rmc})\,\Delta E$ now contributes
to each bin defined by Eq.~(\ref{ebin}), of course.

All previous \lq\lq statistical\rq\rq\ work, up to and including \cite{Dzuba2012}, assumed that the low-energy DR could
be described just in terms of the dielectronic capture, i.e., the fluorescence yield was taken to be unity.
Above $\sim 2$~eV ($\sim 1$~eV) in the DR of W$^{18+}$ (W$^{20+}$) autoionization into the first
excited fine-structure level of the ground term opens-up. Above $\sim 4-5$~eV autoionization
into the first excited term opens-up. In \cite{Badnell2012} we showed that our intermediate coupling
DR cross sections were greatly damped as autoionization into excited states turned-on.
Likewise, the experimental cross section. Recently, Dzuba et al ~\cite{Dzuba2013} applied non-unit
fluorescence yields in their sub-configuration average representation of statistical theory
and they modeled the rapid fall-off of experiment as well. We did not apply our non-unit fluorescence yields
to our partitioned results then. We do so now.

For the present \lq\lq partitioned \& damped\rq\rq\ (PD) approach we apply Eq.~(\ref{red}) to the
total autoionizing width (i.e., with $\nu \rightarrow m$) for use in (\ref{pdri}). On inspection of (\ref{red}),
the autoionization widths are recomputed at each partitioned energy so as to take account of
the closing-off/opening-up at lower/higher redistributed bin energies. We use the
radiative rates associated with the autoionizing levels into which we initially dielectronic capture.
We looked at redistributing over multiply excited (configuration average) states and then using the
radiative rates associated with those states, but we find little sensitivity to the choice. Given that we
actually have a quasi-continuum of chaotically-mixed levels which radiate, either choice seems equally valid.
Using the partitioned bin energy approach we are not restricted in energy, by having to describe all
possible multiply excited autoionizing states, everything is self-contained within the original (two-step)
DR calculation.

The theoretical merged-beam recombination rate coefficient is obtained by convoluting the theoretical cross section
with a flattened Maxwellian electron velocity distribution \cite{Kilgus1992} with the temperatures $k_BT_\|=0.2$~meV
and $k_BT_\perp=20$~meV (section~\ref{sec:exp}).
The TSR dipole magnets field ionize the weakly bound, high-$n$ Rydberg levels of the recombined W$^{17+}$ ion before
they can be detected. The critical principal quantum number for field ionization in this experiment is
$n_{\mathrm{max}} = 68$ \cite{Schippers2001c}. This cut-off quantum number was used for all theoretical
merged-beams rate coefficients.

\section{Results} \label{sec:res}

\begin{figure}[b]
\centerline{\includegraphics[width=\figurewidth]{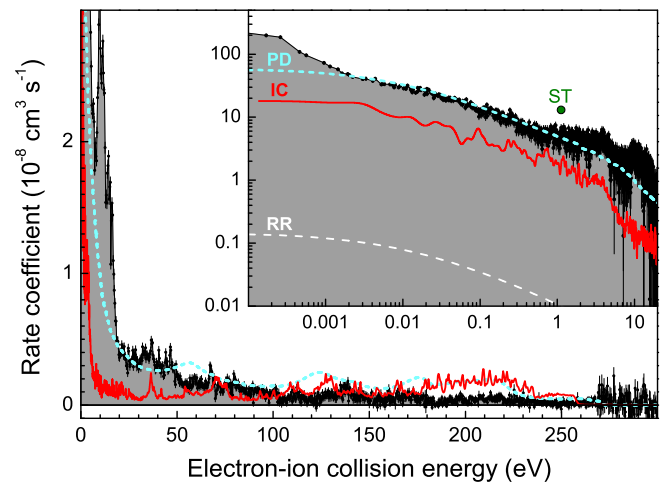}}
\caption{\label{fig:exptheo}(Color online) Comparison of our measured (symbols) and various calculated merged-beam
recombination rate coefficients. The solid curve (labeled IC) is the result of present intermediate-coupling calculation.
The short-dashed curve (labeled PD) is the result of the fully partitioned calculation including autoionizing (and radiative)
damping. The long-dashed curved (labeled RR) is the calculated rate coefficient for radiative recombination. The inset shows the same data up to 20~eV on a double logarithmic scale.
The full circle (labeled ST) is the rate coefficient from the statistical theory by Dzuba \etal~\cite{Dzuba2012}.
}
\end{figure}

\subsection{Merged-beam recombination rate coefficient}\label{sec:resmb}

The measured and calculated merged-beam recombination rate coefficients of W$^{18+}$ are displayed in Fig.~\ref{fig:exptheo} over the energy range 0 to 300~eV. In the collision energy range of 0~eV to about 5~eV the rate coefficient decreases from a value of $\alpha_0 = 2.16 \times 10^{-6}$~cm$^3$~s$^{-1}$ by approximately two orders of magnitude. At higher energies, almost up to the end of the experimental energy range, broad resonance structures are visible. Since their widths are larger than the experimental energy spread, these features are most likely blends of unresolved resonances. The rise of the measured rate coefficient at energies below $\sim 2$~meV is likely caused by additional capture and radiative stabilization of electrons in the time dependent electric and magnetic fields seen by the highly charged ions in their rest frame when travelling through the electron cooler \cite{Gwinner2000,Hoerndl2005c}. These effects are only relevant at very low electron energies. They are disregarded in the comparisons with the present theoretical calculations. The low-energy rise of the experimental merged-beam rate coefficient is also excluded from the experimentally derived plasma rate coefficient where, however, its contribution would be negligible already at electron temperatures much lower than those relevant for fusion plasmas.

Up to at least 1~eV, the calculated RR rate coefficient is always two orders of magnitude smaller than the experimental data. This indicates that the measured rate coefficient is dominated by strong contributions from resonant processes. At low collision energies of up to about 50~eV, the IC results underestimate the measured rate coefficient as well. For electron-ion collision energies between 2~meV and 1~eV, a discrepancy of a factor of 2 to 3 is found. Due to strong resonances which are not reproduced by the IC calculations, the discrepancy between these theoretical results and experimental findings for energies of up to about 50~eV is large. In the collision energy range of 50~eV to 180~eV, IC theory and experiment are in better agreement although there are significant differences in the details of the resonance structures.

Above 180~eV to about 260~eV, the IC theoretical predictions are larger than the results of the measurements whose variations remain below $5 \times 10^{-10}$~cm$^3$~s$^{-1}$. The dominant contribution in the 180~eV to 230~eV range is from $4d$ promotions to $4f$ and $5f$ but here the associated DR resonances can start to autoionize to the $4d^{10}4f^95d$ continuum. As discussed in Ref.~\cite{Badnell2012}, we could not include the $n=5$ continuum due to computational limitations. Likely, what we see by comparison with experiment is the effect of the omission of these suppressed channels. The dominant contribution in the 230 to 260~eV range is from $4f$ promotions to $5l$. They too can access the $n=5$ continuum which has been omitted. But, their contribution is small. Towards the end of the experimental energy range both theory and experiment do not exhibit any significant  contributions from resonant processes to the recombination rate coefficient.

The fully partitioned theory compensates for the limited number of states which were included in the IC calculations, as described in Sec.~\ref{sec:theo}. With damping included in this approach, the absolute rate coefficients from partitioned theory and experiment agree excellently with one another for energies ranging from 2~meV to 1~eV. The shapes of the theoretical and experimental cross section curves in this energy range are nearly identical. At higher energies there are differences in resonance structure but the overall agreement is as good as in case of the IC calculation. The partitioned results are the maximal (damped) ones. Above about 50~eV they are larger than both the experimental and IC results. Above about 180~eV the partitioned results come into agreement with the IC ones as we move to a regime ($A_a<A_r$) where the DR cross sections themselves are largely redistributed without any enhancement.

The result of the statistical theory without damping by Dzuba \etal~\cite{Dzuba2012} is $\alpha = 1.5 \times 10^{-7}$~cm$^3$~s$^{-1}$ for the W$^{18+}$ recombination rate coefficient at an electron-ion collision energy of 1~eV (data point labelled ST in the inset of Fig.~\ref{fig:exptheo}). This value is about three times higher than the experimental rate coefficient at that point. Later Dzuba \etal\ have incorporated damping into their theoretical approach as discussed in Sec.~\ref{sec:theo}. So far, corresponding calculations were carried only for electron-ion recombination of Au$^{25+}$ and W$^{20+}$ ions \cite{Dzuba2013}. Results for W$^{18+}$ are not available.

\subsection{Plasma recombination rate coefficient}

\begin{figure}[b]
\centerline{\includegraphics[width=\figurewidth]{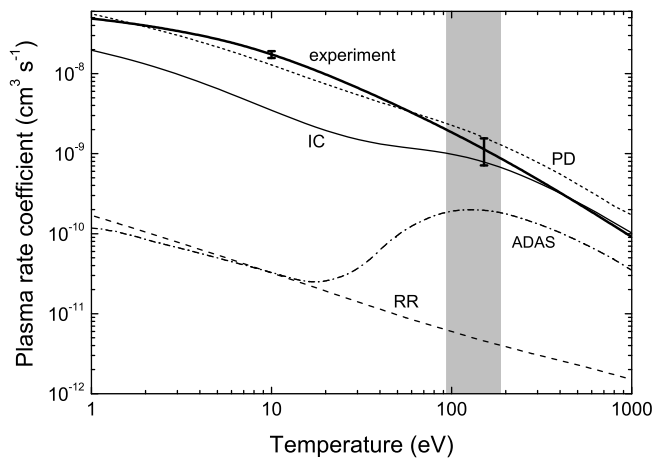}}
\caption{\label{fig:prc}Experimentally derived (thick solid line) and theoretical rate coefficients for electron-ion recombination of W$^{18+}$ in a plasma.
The error bars represent the combined statistical and systematic uncertainty (see text) of the experimentally derived rate coefficient.
The thin solid line (labelled IC) and the dotted line (labelled PD) are the results of present intermediate coupling theory and of the present partitioned-and-damped statistical theory.
The dash-dotted line is the plasma recombination rate coefficient from the ADAS database \cite{Foster2008, ADAS}. The dashed curved is the calculated RR plasma rate coefficient.
The shaded area indicates the plasma temperature range where W$^{18+}$ is expected to form in a collisionally ionized plasma  \cite{Puetterich2005a}.}
\end{figure}

The experimentally derived plasma recombination rate coefficient is obtained from the measured merged-beam recombination rate coefficient essentially by first converting it into a cross section which is then convoluted with an isotropic Maxwellian energy distribution characterized by the plasma electron temperature $T_e$ \cite{Schippers2001c}. Figure \ref{fig:prc} shows the plasma recombination rate coefficient derived from the experimental merged-beam recombination rate coefficient for W$^{18+}$ forming W$^{17+}$, as well as several theoretical results. The plasma temperature range where the abundance of this charge state is expected to peak in a fusion plasma is indicated by the shaded area. At a plasma temperature of 1~eV the experimentally derived rate coefficient is about $5 \times 10^{-8}$~cm$^3$~s$^{-1}$. Towards higher temperatures it decreases monotonically by more than two orders of magnitude over the displayed temperature range. At a temperature above about 250~eV, the present result is to be regarded as a lower limit, since it does not contain any contribution from recombination at electron-ion collision energies above 300~eV. Theoretically, we estimate the missing contribution, from all $n$ above 300~eV and $n>68$ below and from promotions as deep as from $3d$, to be less than 5\% at 1000~eV. This amount decreases rapidly with decreasing temperature until low temperatures where the high-$n$ RR contribution starts to rise again, but it is still no more  than 1\% at 1~eV. The systematic uncertainty of the experimental merged-beam recombination rate coefficient (Sec.~\ref{sec:exp}) leads to a 36\% uncertainty in the plasma rate coefficient around 150 eV. At a 90\% confidence limit, the total relative uncertainty of the experimentally derived rate coefficient, including the missing resonance strength from high-n states, is thus estimated to be $\pm37\%$ at a temperature of 150 eV. In the same way we obtain a total uncertainty of $\pm 10\%$ at a temperature of 10 eV.

To simplify the handling in plasma models, our experimental plasma rate coefficient was fitted in the temperature range 1--1000~eV using
\begin{equation}
	\alpha(T) = T^{-3/2} \sum_{i=1}^6 c_i  \exp\left(-\frac{E_i}{k_BT}\right) \label{eq:fit}
\end{equation}
with $k_B$ denoting the Boltzmann constant. The fit parameters $c_i$ and $E_i$ are given in Tab.~\ref{tab:res}.
In the temperature range 1--1000~eV the fit deviates less than 0.5\% from the experimentally derived plasma rate coefficient.

\begin{table}
\caption{\label{tab:res}Best fit parameters for Eq.~(\ref{eq:fit}), reproducing the experimentally derived plasma recombination
rate coefficient (Fig.~\ref{fig:prc}) with less than 0.5\% relative deviation for temperatures
$1\textrm{~eV} \leq k_BT \leq 1000$~eV. The systematic and statistical uncertainties of the plasma rate coefficient are discussed in the text.}
\begin{ruledtabular}
\begin{tabular}{@{}rdd}
	$i$ & \multicolumn{1}{c}{$c_i$ (cm$^3$ s$^{-1}$ K$^{3/2}$)} & \multicolumn{1}{c}{$E_i$ (eV)} \\
    \hline
	1 & 0.1652	&1.05797 \\
	2 & 0.5085	&4.96985 \\
	3 & 0.8513	&13.7193 \\
	4 & 0.8128	&40.6401 \\
	5 & 0.8851	&102.876 \\
	6 & 0.6247	&232.047 \\
\end{tabular}
\end{ruledtabular}
\end{table}

At a temperature of 1~eV the present IC theoretical result is about a factor of 3 lower than the experimental curve. This deviation decreases at higher temperatures above several 10~eV. In the energy range of interest, i.e., between about 90~eV and 200~eV the IC theory is between 100\% and 25\% lower than experiment. The fully-partitioned-with-damping result agrees better with the experimentally derived rate coefficient, in particular, at temperatures below 100~eV where the deviation is within the experimental uncertainty. The deviation becomes larger at higher temperatures. At 200~eV it amounts to about 43\%.

The DR contribution to the recombination rate coefficient from the ADAS database \cite{ADAS, Foster2008} was calculated using the Burgess General Formula \cite{Burgess1965}. The General Formula is a
high-temperature approximation and contains no description of low-energy DR resonances. At low plasma temperatures, the ADAS rate coefficient is due purely to radiative recombination and so it decreases monotonically up to about 20~eV. In this temperature range it is more than 2 orders of magnitude lower than the experimentally derived plasma rate coefficient.

Resonances lead to the rise of the ADAS rate coefficient at temperatures above 20~eV. The ADAS rate coefficient reaches its maximum at 130~eV where it is a factor of $\sim 7$ lower than the
experimentally derived rate coefficient. This factor varies from 5 to 10 over the temperature range 94--186~eV where  W$^{18+}$ is expected to form in a collisionally ionized plasma \cite{Puetterich2005a}.

\section{Summary and Conclusion} \label{sec:con}

Rate coefficients for the recombination of W$^{18+}$([Kr]$\,4d^{10}\,4f^{10}$) ions with free electrons have been obtained independently on absolute scales from a storage-ring experiment and
from theoretical calculations. Despite adverse experimental conditions, i.e., despite unusually low ion currents and very short beam-storage times, data were obtained with sufficiently low statistical and systematic uncertainty to allow for meaningful comparisons with the theoretical results. The experimental rate coefficient is dominated by particularly strong recombination resonances at very-low electron-ion collision energies below about 10~eV , which also was largely responsible for the short stored ion beam lifetimes seen. These resonances significantly influence the W$^{18+}$ recombination rate coefficient in a plasma even at temperatures of 100--200~eV where W$^{18+}$ is expected to form in a collisionally ionized plasma. These experimental findings for W$^{18+}$ are very similar to the results for recombination of W$^{20+}$ \cite{Schippers2011}.

Our present theoretical IC results for W$^{18+}$ underestimate the experimental rate coefficient by a factor of 2--3 at very low electron-ion collision energies. This is also similar to what has been found for W$^{20+}$ \cite{Badnell2012}. However, the result of our PD statistical theory agrees with the measured rate coefficient excellently for energies of up to about 2~eV, still much better than the IC result at energies of up to 50~eV, and equally well as the IC result at higher energies.

Compared to the W$^{18+}$ recombination rate coefficient from the ADAS database, our experimentally derived rate coefficient in a plasma is more than two orders of magnitude larger for temperatures of up to 10~eV. At higher temperatures, in particular, in the range where W$^{18+}$ is expected to exist in a collisionally ionized plasma, the discrepancy still amounts to factors of 5--10. Since this discrepancy is similar to what has been found earlier already for W$^{20+}$ \cite{Schippers2011} we expect that recombination rate coefficients from the ADAS data base are significantly in error also for tungsten ions of neighboring charge states.

The present fruitful interplay between experiment and theory has clearly lead to a much better understanding of recombination in multi-electron ions with very complex atomic structure. In the near future we will further explore the validity of the theoretical methods by considering neighboring charge states of the tungsten isonuclear sequence. Experimental results for W$^{19+}$  and W$^{21+}$ are currently being analyzed \cite{Krantz2014} with W$^{21+}$, due to its half open $4f$-shell, being a particular challenge for theory.\\

\section*{ACKNOWLEDGMENTS}

We thank the MPIK accelerator and TSR crews for their excellent support.
Financial support by the Deutsche Forschungsgemeinschaft (DFG, contract numbers Mu1068/20-1 and Schi378/9-1) and the
Max-Planck Society is gratefully acknowledged.
MH, ON, and DWS were financially supported in part by the NASA Astronomy and Physics Research and Analysis program
and the NASA Solar Heliospheric Physics program.
NRB was supported in part by the EPSRC (UK) grant EP/L021803/1 to the University of Strathclyde.

%

\end{document}